\documentclass[aps,prl,preprint,amsmath,amssymb,endfloats,12pt,longbibliography]{revtex4-1}
\pdfoutput=1
\usepackage [utf8]{inputenc}
\usepackage{graphicx}% Include figure files
\usepackage{dcolumn}% Align table columns on decimal point
\usepackage{bm}% bold math
\usepackage{upgreek}
\usepackage{url}
\newcommand{\degree}{\ensuremath{^\circ}}

\newcommand{\figsizeb}{12.5cm}
\newcommand{\figsize}{8.5cm}
\DeclareUnicodeCharacter{0301}{\"{u}}

\begin{document}
\title{Analysis of nanometre sized aligned conical pores using SAXS} 
\author{A. Hadley}
\email[]{andrea.hadley@anu.edu.au}
\affiliation{Department of Electronic Materials Engineering, Research School of Physics, The Australian National University, Canberra ACT 2601, Australia.}

\author{C. Notthoff}
\affiliation{Department of Electronic Materials Engineering, Research School of Physics, The Australian National University, Canberra ACT 2601, Australia.}

\author{ P. Mota-Santiago}
\affiliation{Department of Electronic Materials Engineering, Research School of Physics, The Australian National University, Canberra ACT 2601, Australia.}

\author{S. Dutt}
\affiliation{Department of Electronic Materials Engineering, Research School of Physics, The Australian National University, Canberra ACT 2601, Australia.}

\author{S. Mudie}
\affiliation{Australian Synchrotron, Australian Nuclear Science and Technology Organisation, 800 Blackburn Rd Clayton VIC 3168, Australia.}

\author{M. A. Carrillo-Solano}
\affiliation{GSI Helmholzzentrum f{\"u}r Schwerionenforschung, Darmstadt 64291, Germany.}

\author{M. E. Toimil-Molares}
\affiliation{GSI Helmholzzentrum f{\"u}r Schwerionenforschung, Darmstadt 64291, Germany.}

\author{C. Trautmann}
\affiliation{GSI Helmholzzentrum f{\"u}r Schwerionenforschung, Darmstadt 64291, Germany}
\affiliation{Technische Universit{\"a}t Darmstadt, Darmstadt 64289, Germany.}

\author{P. Kluth}
\affiliation{Department of Electronic Materials Engineering, Research School of Physics, The Australian National University, Canberra ACT 2601, Australia.   }

\date{\today}

\begin{abstract}
Small angle x-ray scattering (SAXS) was used to quantitatively study the morphology of aligned, mono-disperse conical etched ion tracks in thin films of amorphous SiO${_2}$ with aspect ratios of around 6:1, and in polycarbonate foils with aspect ratios of around 1000:1. This paper presents the measurement procedure and methods developed for the analysis of the scattering images and shows results obtained for the two material systems. To enable accurate parameter extraction from the data collected from conical scattering objects a model fitting the two dimensional (2-D) detector images was developed. The analysis involved fitting images from a sequence of measurements with different sample tilts to minimise errors which may have been introduced due to the experimental set up. The model was validated by the exploitation of the geometric relationship between the sample tilt angle and the cone opening angle, to an angle observed in the features of the SAXS images. We also demonstrate that a fitting procedure for 1-D data extracted from the scattering images using a hard cylinder model can also be used to extract the cone size. The application of these techniques enables us to reconstruct the cone morphologies with unprecedented precision.
\end{abstract}

% insert suggested PACS numbers in braces on next line

\maketitle %\maketitle must follow title, authors, abstract and \pacs
%%%%%%%%%%%%%%%%%%%%%%%%%%%%%%%%%%%%%%%%%%%%%%%%%%%%%%%%%%%%%%%%%%%%%%%%%%%%%%%%%%%%%%%%%
\section{Introduction}

Nano-scale pores in cellular membranes have the ability to sense and control the passage of ions through the cell walls. The capability to artificially mimic this behaviour is highly desirable since it forms the basis for chemical- and bio-sensing at the molecular level. The ion track etching method can be used to create artificial uniform and parallel narrow channels with controlled openings as small as just a few nanometres \cite{Apel2001a, Siwy2006} in a variety of materials including polymers and ionic crystals \cite{Fleischer1975, Trautmann2009, Trautmann1996, Ferain2000}, various glasses, minerals and inorganic insulators \cite{Siwy2006, Vlassiouk2009, Afra2013, Dallanora2008, Dekker2007, Hadley2019, Kaniukov2016}. Ion track etched pores in polymers are extensively used in commercial applications such as filtration and laboratory cell culture (for example, see \cite{Sterlitech2019}) and as sensors \cite{Han2011, Perez2017, Perez2018} and templates for nanowire growth \cite{Toimil2012}. Artificial solid state pores fabricated in inorganic materials such as SiO$_2$ and Si$_3$N$_4$ are very interesting for sensing applications since they are chemically and thermally stable, and have the potential to be able to be integrated with standard electronic device processing \cite{Dekker2007}.

Irradiation with swift heavy ions induces damage along the ion's trajectory that is confined to a region of around 5 nm radially along the track. The resulting cylindrical ion tracks can be up to tens of micrometres long, depending on the ion energy and the thickness of the irradiated material. Each individual ion produces an ion track. The number of tracks is thus controlled by the number of ions the material is exposed to. In most materials the damage created along the ion path can be preferentially dissolved by chemical etching converting the ion track into an open channel. These pores have the potential to be tailored for specific applications by controlling the pore shape and number density, as well as their surface chemistry. However, to realise these promising applications a better understanding of the ion track etching process on the initial track damage is needed.

Many studies have been conducted on etched ion tracks in polymers whereas information on track etching in SiO$_2$ is somewhat limited \cite{Dallanora2008, Kaniukov2016, Jensen2006, Vlasukova2014}, and almost exclusively for the purpose of identifying tracks in this material and determination of the track formation threshold. Systematic studies of the resulting etched pore morphology in SiO$_2$ and its dependence upon irradiation and etching conditions are needed to reliably reproduce ion track etched pores tailored for specific applications. Typically their geometry and size are estimated from cross section scanning electron microscopy (SEM) images \cite{Apel2001a, Apel2014, Perez2018a} or from electrodeposited metal replicas \cite{Duan2016, Karim2009} and can be very difficult to characterise accurately. Our study addresses these issues by using small-angle x-ray scattering (SAXS) as the main method for characterising the etched ion track channels. SAXS has previously been used to characterize unetched cylindrical ion tracks \cite{Bierschenk2013, Engel2009,  Kluth2008a, Kluth2008b, Kuttich2013} and to a lesser extent etched ion tracks in SiO$_2$ and the mineral apatite \cite{Hadley2019, Cornelius2010, Nadzri2017}, but an appropriate model for SAXS of conical etched ion tracks has not been realised. In this paper we present the analytical techniques we developed for accurately reconstructing the size and shapes of conical etched pores and apply them to two different material systems. The first example is the analysis of SAXS data from etched conical ion tracks in SiO$_2$ with aspect ratios around 6 to 1. The second example is the application to very high aspect ratio (around 1000 to 1) narrow conical etched channels in polycarbonate (PC) foil membranes. Application of the analysis techniques to these two very different material systems illustrates the respective advantages of the analytical approaches which were developed.

The intensity pattern resulting from a scattering experiment contains information about the morphology and size distribution of the etched ion tracks. In contrast to most other types of structures measured by SAXS (such as biological molecules, proteins etc.) the ion track etched nanopores are well aligned with each other and have very narrow size distributions, as they result from the etching of ion tracks generated by monoenergetic ions \footnote{the relative energy precision is estimated to be better than 0.1\% {\citep{Skorka1981, Lobanov2018}}} with well-defined ion beam direction \footnote{The maximum beam deflection due to the geometry of our experimental set-up is estimated to be less than 0.08\degree}. The uniformity of etched ion tracks has been confirmed in many previous studies \citep{Dallanora2008, Jensen2006}. Toimil-Molares {\textit{et al.} \cite{Toimil2001} confirmed the uniformity and alignment of etched ion tracks in polycarbonate by performing x-ray diffraction (XRD) on electrodeposited nanowires inside an ion etched template to asses the alignment of the pores with each other within the template. The XRD independently confirmed the high degree of alignment of the pores.  

The polydispersity of the track radii for cylindrical ion tracks has previously been modelled and for ion tracks in amorphous SiO$_2$ irradiated with 185 MeV Au ions has been found experimentally to be less than 2\% \citep{Kluth2008b}.  With an appropriate geometric model for the etched tracks, we can thus determine the average size parameters from the scattering image recorded by the detector with high precision. Once data has been collected by SAXS, several steps must be followed to extract the parameters which describe the morphology of the structures being measured. This procedure usually involves data reduction, background subtraction, and fitting of the scattering images. 

In previously published work \cite{Hadley2019} a systematic study of the dependence upon irradiation and etching conditions of ion track etching in a-SiO$_2$ was conducted. In \cite{Hadley2019} the SAXS technique combined with chemical etching allowed us to determine ion track etch rates in both the axial and radial directions, as well as the cone opening angles, for a range of ion energies and etchant concentrations. Data was analyzed using an average of the results obtained for a tilt sequence of measurements for each sample from an ion energy/etchant concentration series. In the present work we show in detail how this analysis technique was applied and extend it by developing three distinct analysis techniques to quantitatively analyze the size and shape of the conical etched channels. Briefly, the three analysis techniques are; (i) 2-D fit: Opening angle, length and radii of the cones are determined by the application of a 2-D fitting process on the anisotropic SAXS images recorded by the detector. The 2-D fitting routine allows simultaneous fits of two images to determine the sample tilt angles accurately. (ii) 1-D data fit: Fit of 1-D detector image slices to determine the cone dimensions, and (iii) Angle fit function: a geometric relationship between the cone opening angle, the sample tilt angle and the angle between features in the scattering images was developed to determine the cone angles and any offset in the tilt angles directly from the detected scattering images without the need to analyze the intensities.

\begin{figure}
 \includegraphics[width=\figsize]%
 {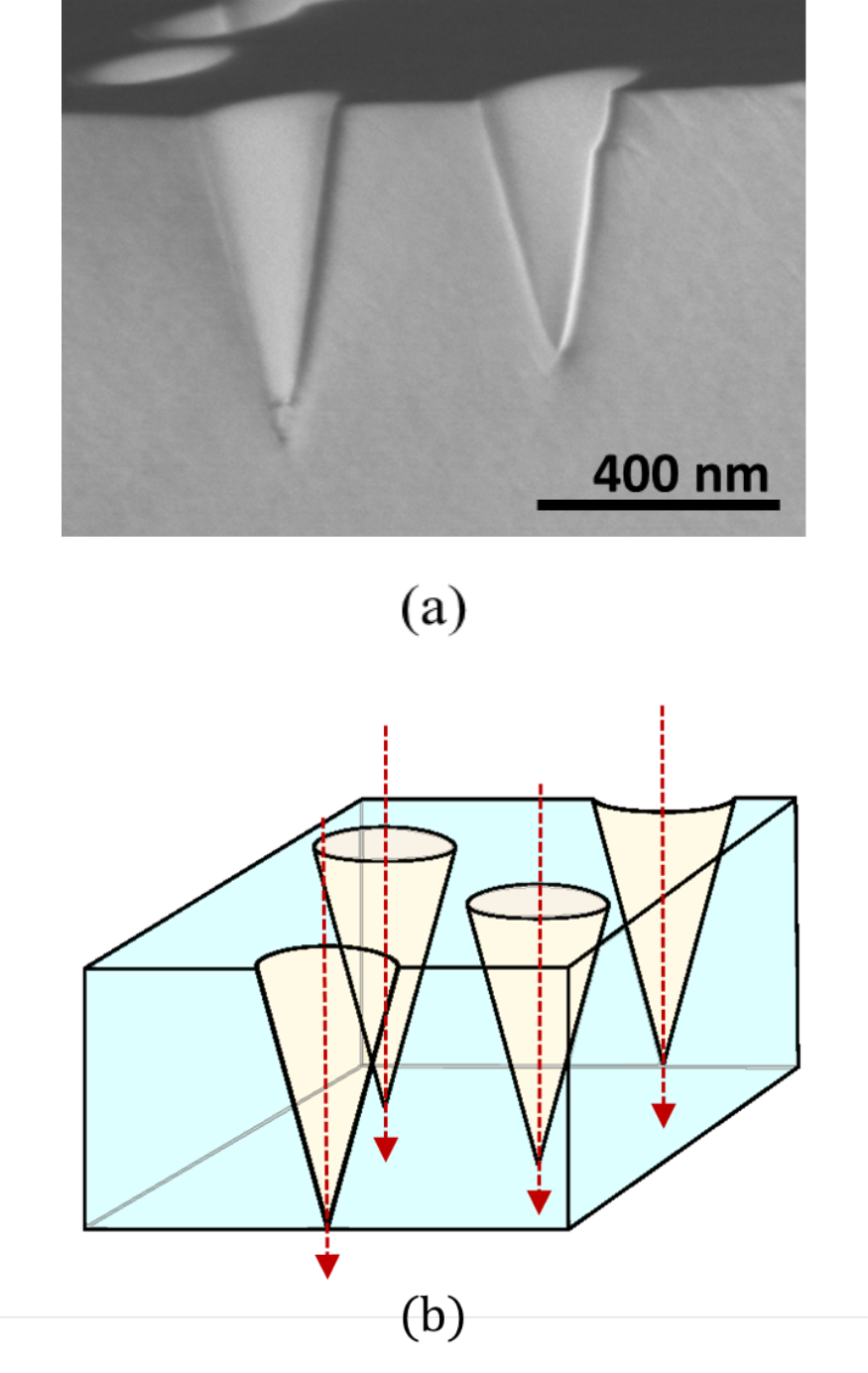}
 \caption{(a) Cross-section SEM image of conical etched ion tracks in a-SiO$_2$, (b) Schematic diagram of etched cones including a cross-section of a cone, showing the paths of the ions in red.}
  \label{fig:Fig1}
\end{figure}

The general shape of the etched channels in SiO$_2$ is conical as shown by SEM micrograph in Fig.~\ref{fig:Fig1}(a) and by the schematic view in Fig.~\ref{fig:Fig1}(b). In Fig.~\ref{fig:Fig1}(a) the sample has been cleaved through the centre of two cones to reveal close to their full depth. The cone on the right is slightly set back from the cleaved edge and has not cleaved through its centre. The openings of two more cones on the upper surface of the sample can be seen at the upper left of the image. In Fig.~\ref{fig:Fig1}(b) the paths of the normally incident swift heavy ions along the cone axes are indicated in red. SEM imaging was used to confirm the general shape of the etched pores, thereby informing our choice of analytic model. The general shape of the high aspect ratio conical etched pores in polycarbonate membranes can be seen in the SEM images shown in \cite{Perez2018, Duan2016}. Reference \cite{Duan2016} shows electrodeposited Cu replicas formed inside conical channels etched under similar conditions to some of the samples analyzed in the present work. 

The studies performed so far on etched ion tracks in SiO$_2$ used SEM, transmission electron microscopy (TEM) or atomic force microscopy (AFM) as the primary means for determining the etched pore morphologies \cite{Dallanora2008, Kaniukov2016, Jensen2006, Vlasukova2014}. While these techniques are useful, they can potentially introduce measurement artefacts due to sample preparation, generally suffer from poor statistics due to the limited number of channels imaged and have limited accuracy in quantifying the pore dimensions. SAXS is a powerful tool for the characterisation of ion tracks \cite{Kluth2008a} and nano-sized channels \cite{Hadley2019, Engel2009, Hossain2017, Pepy2007} in many materials.  It has several advantages over other characterisation techniques such as electron microscopy. The SAXS technique provides superior statistics since scattering from many structures is measured at once. For an x-ray beam spot size of 25 x 100 $\upmu$m$^2$ and an ion fluence of 1 x 10$^8$ ions cm$^{-2}$, approximately 2.5 x 10$^5$ structures are measured simultaneously. The results so deduced are an excellent statistical representation of the whole population. SAXS is a non-destructive technique that does not introduce any measurement artefacts due to sample preparation and measures the entire pore volume. These attributes are particularly advantageous for high aspect ratio nanopores, as not only is it possible to resolve the small changes in the pore sizes across the sample population, but changes along the pore length can also be resolved. Data acquisition times are short (in the order of 1 to 10 seconds) and the samples are measured in air so it is possible to conduct a variety of time-resolved \textit{in situ} studies (for examples see \cite{Afra2011, Schauries2013}).

%%%%%%%%%%%%%%%%%%%%%%%%%%%%%%%%%%%%%%%%%%%%%%%%%%%%%%%%%%%%%%%%%%%%%%%%%%%%%%%%%%%%%%%%%
\section{Experimental}
%%%%%%%%%%%%%%%%%%%%%%%%%%%%%%%%%%%%%%%%%%%%%%%%%%%%%%%%%%%%%%%%%%%%%%%%%%%%%%%%%%%%%%%%%
   \subsection{SiO$_2$ samples}
   
SiO$_2$ samples were prepared from 2 $\upmu$m thick thermally grown thin films  using swift heavy ion irradiation followed by chemical etching as described in \cite{Hadley2019}. Irradiation with 54, 89, 185 MeV (ANU Heavy Ion Accelerator Facility) and 1.1 GeV (UNILAC linear accelerator of the GSI Helmholz Centre) Au ions was carried out at room temperature normal to the sample surface with applied ion fluences between 1x10$^{8}$ to 1x10$^{9}$ ions cm$^{-2}$ to avoid significant overlap of the etched pores. The samples were subsequently chemically etched in dilute (2.5\% or 5\% by volume) HF, yielding parallel, almost mono-disperse conical etched channels with their axes along the trajectory of the ion tracks as shown in Fig.~\ref{fig:Fig1}(b). SEM images were used to directly confirm the general morphology of the etched channels was conical, and data representative of the measured samples was chosen to illustrate the analysis techniques which were developed to characterise the etched cones and applied in \cite{Hadley2019}.

   \subsection{Polycarbonate samples}
   
Polycarbonate (PC) foils (Makrofol N, Bayer AG, 30 $\upmu$m thick) were irradiated with 2.2 GeV Au ions at the GSI UNILAC accelerator with a fluence of 10$^6$ ions cm$^{-2}$. The samples were etched in a solution of 9M sodium hydroxide (NaOH) and methanol at a temperature of 50\degree{} C for a range of etching times from 7 to 80 minutes. To produce conically shaped pores, asymmetric etching was performed by placing the etchant on one side of the irradiated PC foil and water on the other, as described in \cite{Karim2009}. The current flow across the membrane was monitored and etching was stopped by flushing with water after an increase in current flow was detected, indicating that the pores had etched through the PC foil. It has previously been shown that the addition of methanol to NaOH affects the ratio of the bulk etch rate to the track etch rate in PC, resulting in conically shaped etched channels \cite{Duan2016, Karim2009}. A series of samples was etched with varying methanol concentrations, from 0 to 95 volume per cent and the effect of methanol concentration on the cone opening angles was studied by SAXS.

   \subsection{SAXS measurements}
   
SAXS measurements were performed at the SAXS/WAXS beamline of the Australian Synchrotron in Melbourne, Australia. They were performed in transmission mode with a photon energy of 12 keV and a sample to detector length of approximately 7200 mm. The images were recorded using a two dimensional (2-D) Pilatus 1M hybrid pixel detector using exposure times between 0.5 and 5 seconds. Accurate calibration of the q range was done for each group of measurements using a Silver Behenate (AgBeh) standard following the instructions given in \cite{Scatterbrain2019}. Calibration is based on the first order AgBeh ring. At the long camera length of approximately 7200 mm and an x-ray photon energy of 12 keV the q range is 0.002 to 0.100 \AA$^{-1}$.  The detector is mounted on a horizontal and vertical motorized stage. To avoid gaps in the images resulting from the gaps between detector modules, three images are obtained by an automated system which moves the detector between three exposure positions and assembles a composite image. Specific details of the Pilatus 2-1M detector are given in \citep{Kirby2013}.

Samples were mounted on a three axis goniometer, as shown in Fig.~\ref{fig:Fig2}, to enable alignment of the etched channels with the x-ray beam. The three directions of rotation of the goniometer motors, Pitch around the X axis, Yaw around the Y axis and Roll around the Z axis, are shown in the schematic at the right of the figure. The Z axis is parallel to the direction of the incident x-ray beam. The image on the left of Fig.~\ref{fig:Fig2} shows the sample stage with the sample holder just below the position of the x-ray beam. The sample is mounted on a holder with a separate small rotating motor behind it (not shown), allowing 360\degree{} Pitch rotation about the X axis. The inset below is a close-up view of the sample holder which is almost vertical in the image.

In practice, the alignment is done by adjusting the position of the sample in the Pitch and Yaw directions until a uniform set of concentric rings appears in the scattering pattern. This indicates that the pore axis is close to parallel to the direction of the x-ray beam. A tilt sequence of measurements is then performed by adjusting the Pitch motor by an angle $\gamma$ from 0\degree{} to 50\degree{} in 5\degree{} increments (or 1\degree{} increments for selected samples). The relative accuracy of the tilt angle with respect to the initial alignment position is accurate to better than 0.25\degree. The $\gamma$ tilt sequence enables us to determine if any offset in the initial alignment of the sample was introduced, therefore allowing more accurate determination of the extracted parameters.

\begin{figure}
 \includegraphics[width=\figsize]%
 {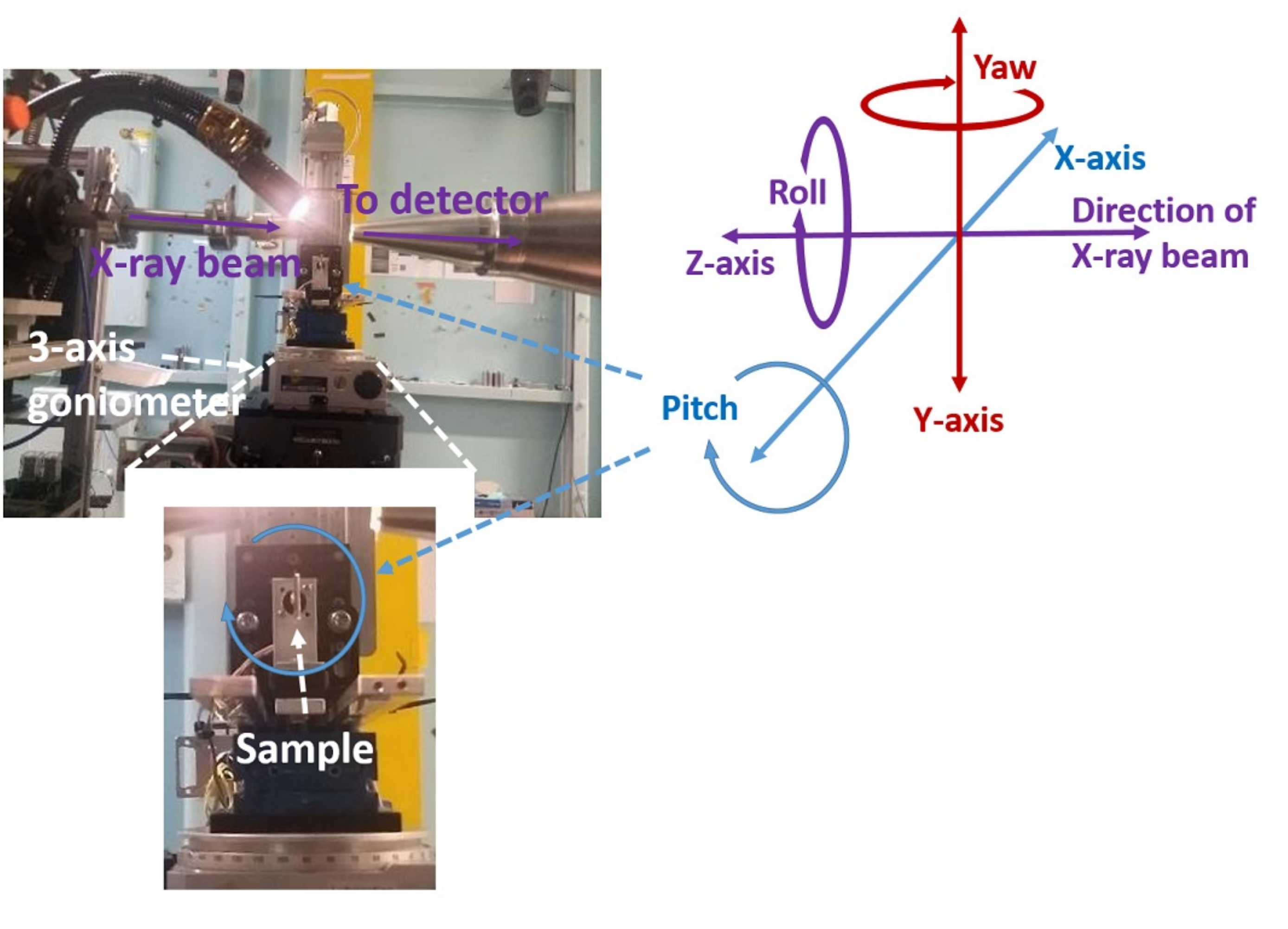}
 \caption{Photographs of sample positioning in the SAXS/WAXS beamline at the Australian Synchrotron. The photograph on the left shows the rotating sample holder, with an enlarged view below. The three axes of rotation are shown in schematic view on the right.} 
  \label{fig:Fig2}%
\end{figure}

To account for the effect of background scattering from air, an air shot (no sample in the path of the x-ray beam) which is taken during calibration was subtracted when analyzing the 1-D data. When using the 2-D fitting routine to fit the 2-D data a constant background is included as one of the fitting parameters.

\subsection{SAXS image analysis}

Figure~\ref{fig:Fig3}(a) shows a schematic of a conically shaped scattering object coaxially aligned with the direction of the x-ray beam. The scattering intensity pattern obtained resembles that of a pin-hole scattering experiment, appearing as a series of concentric rings related to the dimensions of the real space object by its Fourier transform, as shown by the example experimental image to the right of Fig. \ref{fig:Fig3}(a). When the cones are tilted by an angle $\gamma$ with respect to the x-ray beam as depicted in the schematic diagram in Fig. \ref{fig:Fig3}(b), the scattering intensity pattern develops two sets of streaks, as shown by the example experimental image given to the right of Fig. \ref{fig:Fig3}(b). Due to the rotational symmetry of the conical pores, a cylindrical coordinate system fixed on the axis of a radially symmetric cone of height H and base radius R$_0$ is chosen. For a cone of height H, H/4 indicates the position of the cone centroid which was arbitrarily chosen as the pivot point of the cones in Fig. \ref{fig:Fig3}(b). The choice of pivot point does not affect the result, since the amplitude of the Fourier transform is translationally invariant \citep{Sivia2011}. The angle between the streaks in the scattering patterns from the tilted samples is related to the half cone opening angle $\beta$ and the tilt angle $\gamma$, while the intensity oscillations along the streaks contain information related to the cone size over the entire length of the cone. Our model assumes a monodisperse system with randomly distributed and well separated scattering objects. The fact that the model fits the experimental data very well indicates that the size distributions of R$_0$, H and $\beta$ are very low and that there is no order between the etched cones.

\begin{figure}
 \includegraphics[width=\figsize]%
 {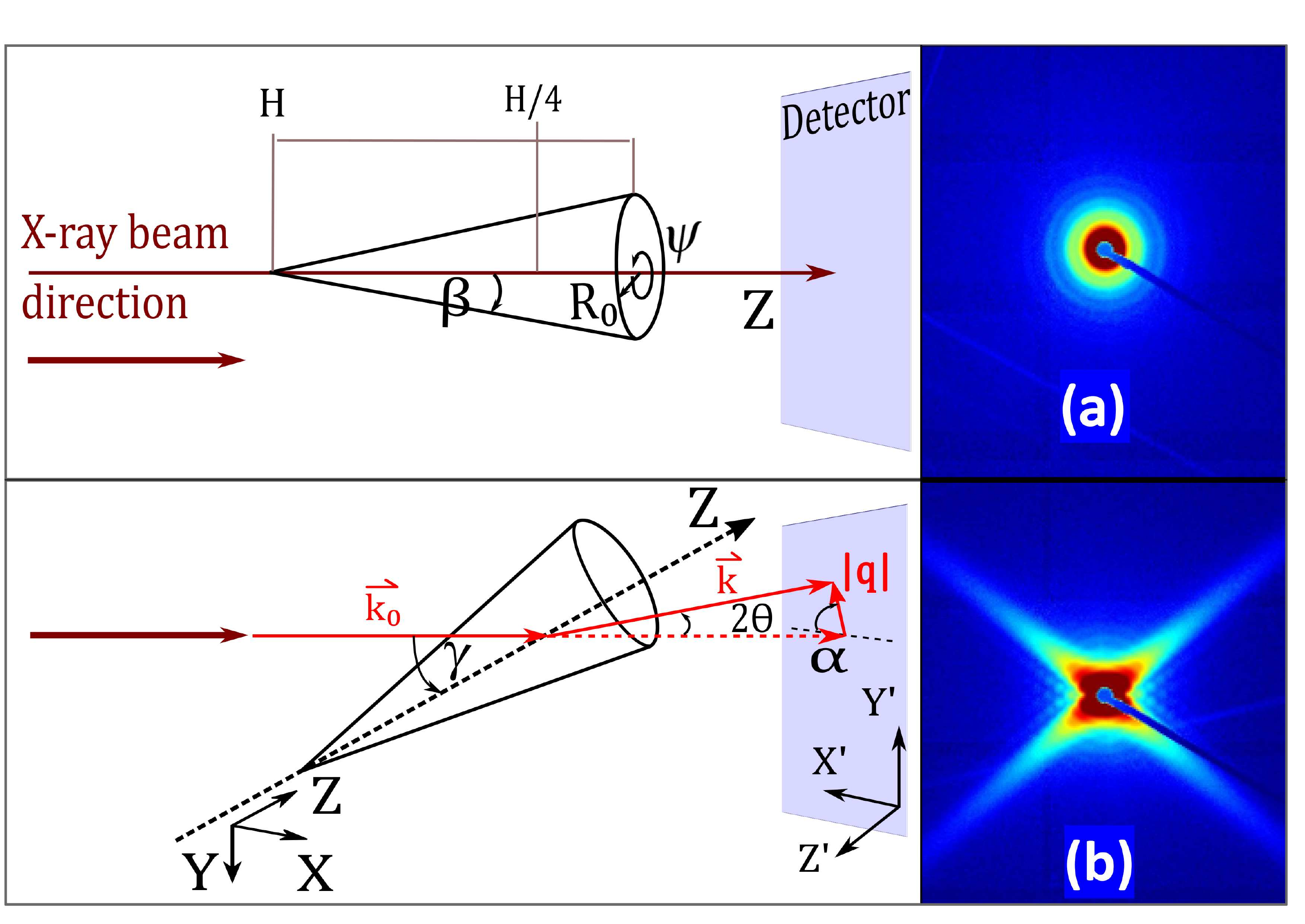}
 \caption{(a) Experimental configuration for conical pores coaxially aligned with the direction of the x-ray beam, and example of resulting experimentally measured isotropic scattering image. (b) Experimental configuration for conical pores tilted by an angle $\gamma$ with respect to the direction of the x-ray beam, and example of resulting experimentally measured non-isotropic scattering image.}
  \label{fig:Fig3}%
\end{figure}

To analyze the data, a step-wise approach was adopted. Both 1-D and 2-D fits were performed to allow a comparison of both methods; 

(i)	The conical shape of the etched ion tracks was confirmed by SEM. An example is shown in Fig. \ref{fig:Fig1}(a).

(ii)	A model for conically shaped scattering objects was developed and implemented as a numerical least squares fit of the 2-D scattering images. This method provides accurate information on the cone length, radius and opening angle.

(iii)	1-D data was extracted and fit using a hard cylinder model, revealing accurate information on the cone radii.

(iv)	Relationship between $\gamma$, $\beta$, and the angle between the streaks in the scattering image $\phi$. This method provides an independent and accurate way to measure $\beta$ without needing to analyze the intensities in the scattering images.

\subsubsection{Cone model}

Interference from x-ray scattering occurs wherever there is a change in electron density in the material being studied. The detected interference pattern contains information about the size and shape of the scattering objects. The scattering vector \textbf{q} is given by 
\begin{equation}
\textbf{q} = \textbf{k$_0$} \; - \; \textbf{k} %
\label{eqn:Scq},
\end{equation} 
where \textbf{k$_0$} is the incident x-ray wave vector and \textbf{k} is the wave vector of the scattered photons. Mathematically the scattering amplitude, which is related to the form factor, is a Fourier transform of the three-dimensional variation in electron density, or volume of the scattering objects. For a three-dimensional real space scattering object, the general expression for the scattering amplitude is
\begin{equation}
f(\textbf{q}) = \iiint \; \rho(\textbf{r}) \, \mathrm{exp}(-i(\textbf{q.r})) \, d^{3}r %
\label{eqn:eqn1}.
\end{equation} 

For monodisperse and parallel conical etched ion tracks a reference frame co-axial with the axis of the cones is chosen as shown by the dotted line labelled Z and the angle of rotation about the Z axis $\psi$  in Fig.~\ref{fig:Fig3}(a) and (b). The scattering amplitude may then be given as
\begin{equation}
f(\textbf{q}) = \int^\infty_{-\infty} \int^\pi_{-\pi} \int^\infty_{-\infty} \; \rho(r,z)\; \mathrm{exp}(-i(\textbf{q.r}))r\,dr\; d\psi \; dz
\label{eqn:eqn2},
\end{equation}
where $\rho$ is the electron density of the material. We assume that the density inside the etched cones is constant. This assumption is realistic since the conical structures are hollow after track etching. For conical scattering objects of height H, 
\begin{equation}
\rho = \biggr\lbrace ^{\Delta\rho, \;\; for \; r \; \leq \; r_z \; and \;0 \, \leq \, z \, \leq \, H}_{0 \;\;\;\; otherwise} 
\label{eqn:eqn3},
\end{equation}
where r$_z$ is the value of the radius at any position along the Z axis and may be calculated by
\begin{equation}
r_{z} = z \, \mathrm{tan} \, \beta,
\label{eqn:eqn4}
\end{equation}
where $\beta$ is the half cone opening angle and $\Delta\rho$ is the density difference between the etched region of the ion track (i.e. density of air) and the surrounding un-etched material. It is assumed that there is an abrupt boundary from the interior of the etched cones to the surrounding material. In the reference frame of the real space cones the radial component of the vectors \textbf{q} and \textbf{r} may be expressed as
\begin{equation}
r =  \sqrt{x^{2} + y^{2}}, \; q_{r} = \sqrt{q_{x}^{2} + q_{y}^{2}}
\label{eqn:sqrr} .
\end{equation}
The product $\textbf{r.q}$ is therefore given by
\begin{equation}
\textbf{r.q}=r q_{r} \mathrm{cos}\varphi + q_{z}z
  \label{eqn:eqn7} ,
\end{equation}
where $\varphi$ is the azimuthal angle between $\textbf{r}$ and $\textbf{q}$.
The scattering amplitude may now be given by
\begin{equation}
f(\textbf{q}) = \int^H_{0}  \int^{r_z}_{0}  \int^{2\pi}_{0} \Delta \rho \,r\, \mathrm{exp}(-i(q_{r}r \mathrm{cos} \varphi + q_{z}z)) d\varphi \, dr \, dz
  \label{eqn:eqn8}.
\end{equation}
Because we have assumed rotational symmetry about the Z axis of the cones, the integral in the angular dimension can be solved to give 
\begin{equation}
f(q_{r},q_{z}) = \int^H_0\int^{r_z}_0 2\pi\Delta\rho \; J_0 (rq_{r})\;r\; \mathrm{exp}(-izq_{z})dr\;dz
\label{eqn:eqn9}
\end{equation}
where $J_0$ is the Bessel function of the zero$^{th}$ order. Evaluating the integral in the radial direction, Equation~\ref{eqn:eqn9} simplifies to 
\begin{equation}
f(q_{r},q_{z}) = 2\pi \Delta\rho \int^H_0 \; J_1 (r_{z} q_{r})\;\frac{r_z}{q_r}\; \mathrm{exp}(-izq_{z})\,dz
\label{eqn:eqn10}
\end{equation}
where J$_1$ is the Bessel function of the first order. This expression for the scattering amplitude for conical scattering objects accounts for the Z dependence of the radius along the length of the cones. 

The reference frame that we choose for our real space objects (the cones) is the axis of the cones, shown in Fig. \ref{fig:Fig3} denoted Z, which is fixed to the cones and moves with them. The experimentally accessible parameters are the scattering angle $\theta$, the magnitude of the scattering vector $\vert \textbf{q} \vert$ and the azimuth angle $\alpha$,  which are recorded on the 2-D detector in the detector or laboratory coordinate system x$^\prime$, y$^\prime$, z$^\prime$. $\alpha$ is defined as positive when rotating in a clockwise direction from the y$^\prime$ axis ($\alpha$ = 0\degree) towards the negative x$^\prime$ axis. A rotation of the sample by an angle $\gamma$ about the x$^\prime$ axis in the lab-coordinate system is equivalent to a rotation of the $\textbf{q}$ vector in the reference frame of the real space cone. Choosing a left-handed coordinate system by convention, equations~\ref{eqn:eqn11}(a) - (c) enable translation between the 2-D reference frame of the detector (Fourier space) and that of the real space objects \cite{Engel2009},
 \begin{subequations}
    \begin{align}
      q_{x} &= -\;|\textbf{q}| \;\mathrm{sin}\alpha\;\mathrm{cos}\theta  \\
      q_{y} &= |\textbf{q}|\;(\mathrm{cos}\alpha\;\mathrm{cos}\gamma\;\mathrm{cos}\theta\;+\;\mathrm{sin}\gamma\;\mathrm{sin}\theta) \\
      q_{z} &= |\textbf{q}|\;(\mathrm{cos}\alpha\;\mathrm{sin}\gamma\;\mathrm{cos}\theta\;-\;\mathrm{cos}\gamma\;\mathrm{sin}\theta) \, . 
    \end{align}
   \label{eqn:eqn11}
  \end{subequations}

We numerically calculate Equation~\ref{eqn:eqn10} and use an open-source Python implementation of a least squares fitting routine (LMFIT \cite{Newville2014}, \footnote{available from http://doi.org/10.5281/zenodo.3588521}, as reported in \cite{Hadley2019}) to fit it to the experimental data. Experimental misalignments of the cone axis in the direction orthogonal to $\gamma$ were implemented by applying a rotation matrix to Equation~\ref{eqn:eqn11} (a)-(c), representing a rotation around the y$^\prime$ axis. 

The radius at the base of the cone, $R_0$ may be calculated using Equation~\ref{eqn:eqn4}
\begin{equation}
R_{0}=H \, \mathrm{tan} \, \beta
  \label{eqn:eqn12}.
\end{equation}

%\footnote{available at \url{https://zenodo.org/record/3381550{\#}.XXIAsnsRUUE}}

\subsubsection{1-D cylinder model}

For the case of cylindrically shaped scattering objects of length L and with a constant radius R, in the limit of $q_{z}\rightarrow0$, Equation~\ref{eqn:eqn10} may be solved analytically to yield 
\begin{equation}
f(q_{r}) = \frac {2\pi\,L\,R\,\Delta\rho}{q_{r}} \; J_1 (Rq_{r})
\label{eqn:eqn13}
\end{equation}
where $J_{1}$ is the Bessel function of the first order \cite{Kluth2008a}. Along the streaks the condition $q_{z}\rightarrow0$ is satisfied. The total scattering intensity I(q$_r$) for N scattering objects is given by
\begin{equation}
I(q_r) \propto N \, \vert \, f(q_{r})\vert^2
\label{eqn:eqn14}.
\end{equation}

For the case of cylindrical etched ion tracks, it is appropriate to model the change of density in the radial direction as an abrupt increase at the interface between the etched region and the remaining un-etched material \cite{Nadzri2017}, as for conically shaped etched tracks. To include slight variations in the density profile of the radii of an actual sample, a narrow Gaussian distribution is convoluted with the radius. The scattering intensity from an ensemble of N etched ion tracks of radius R is then given by
\begin{equation}
I(q_r,R,\sigma) = \frac{N}{\sqrt{2\pi}\sigma} \int^\infty_{-\infty}\vert f(q_{r}) \vert^{2}  \mathrm{exp} \left( \frac{-(r-R)^2}{(2\sigma_r)^2} \right) \; dr
\label{eqn:eqn15} .
\end{equation}

This one dimensional cylinder model has successfully been used to characterise both un-etched ion tracks \citep{Bierschenk2013, Kluth2008b, Afra2011} and cylindrical etched ion tracks \cite{Nadzri2017, Hossain2017} in a variety of materials. For the case of a simple cylinder the total intensity I(q) is the incoherent sum (Equation~\ref{eqn:eqn15}) of the scattering contributions from all of the cylindrical pores being measured simultaneously \cite{Cornelius2010}.

The 1-D cylinder model can be used as an approximation to analyse data from very small or very high aspect ratio cones which show very weak oscillations ($R\leq20$ nm) and are therefore difficult to fit using the 2-D data fitting method described above. To provide an appropriate approximation in the model of the variation of the cone radii along their length, a box shaped distribution is convoluted with the radius values along the length of the conical etch channels using the box distribution function $\Theta$, 
\begin{equation}
\Theta(r, r_0, W) = \biggr\lbrace ^{1, \;\; for \;|r\; - \; r_0| \; \leq \; W}_{0,\;\; otherwise} 
\label{eqn:eqn16},
\end{equation}
where W is the width of the radius distribution and r$_0$ is the average radius. This provides a more realistic approximation of the variation of the cone radii along the length of the cones. The radius value extracted by the 1-D fit corresponds to the volume weighted average radius, confirmed by comparison with the radius values obtained using the 2-D fitting model. As the radius variation along the length of  each cone is more dominant than the radius variation between cones, the distribution is taken as the coherent sum of all the scattering contributions and the scattering intensity is given as
\begin{equation}
I(q_r,r_0,W) \propto \; \biggr\vert \frac{N}{2W} \int^\infty_{-\infty} f(q_{r})  \; \Theta(r, r_0, W) \; dr \biggr\vert^{2}
\label{eqn:eqn17}.
\end{equation}

\subsubsection{Relationship between the angles $\gamma,\, \beta$ and $\phi$}

As is apparent from Fig.~\ref{fig:Fig4}, the angle between the two streaks in the scattering image is dependent on the sample tilt angle $\gamma$. Using the following empirical relationship,
\begin{equation}
\mathrm{tan} \, \beta = \mathrm{tan} \, \phi \, \mathrm{sin} \, \gamma
 \label{eqn:eqn18},
\end{equation}
the cone opening angle $\beta$ may be deduced from a tilt sequence of scattering intensity images, where $\phi$ is the angle between the streaks measured directly from the image as shown in Fig.~\ref{fig:Fig5}. The detector image is the Fourier transform of the physical cones through which the angles are preserved, therefore we can directly determine the value of $\beta$ from any of the scattering images provided we know the exact value of $\gamma$. Equation~\ref{eqn:eqn18} can be derived from the geometrical projection of the cone  dimensions onto a plane perpendicular to the x-ray beam with its origin at the apex of the cone, as shown in the supplementary information. Using this relationship it is possible to determine $\beta$ without the need to analyse the intensity versus scattering vector data which is an advantage, particularly where the intensity values are weak and don't show any oscillations.

\begin{figure}
 \includegraphics[width=1\textwidth]
 {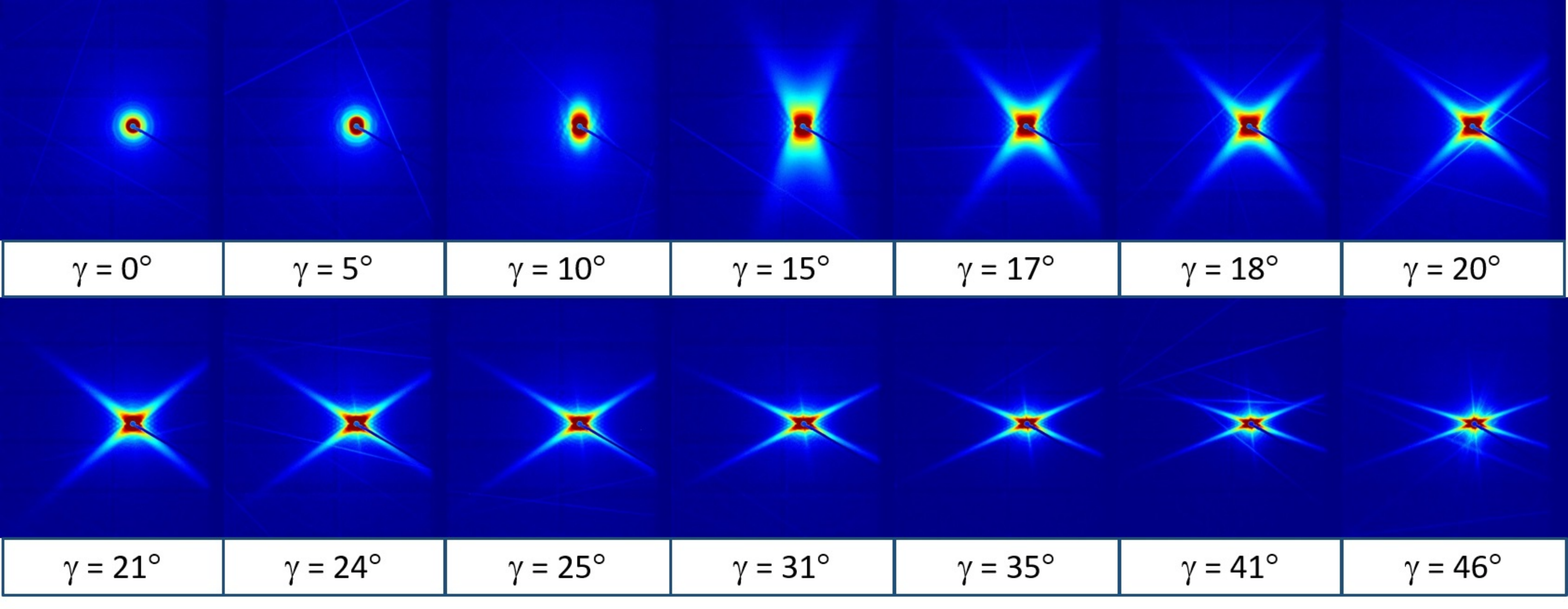}
 \caption{$\gamma$ tilt sequence of experimental detector images for a 2 $\upmu$m a-SiO$_{2}$ sample irradiated with 89 MeV Au ions and etched in 2.5 volume per cent HF for 12 minutes. }
  \label{fig:Fig4}
\end{figure}

\begin{figure}
 \includegraphics[width=\figsize]%
 {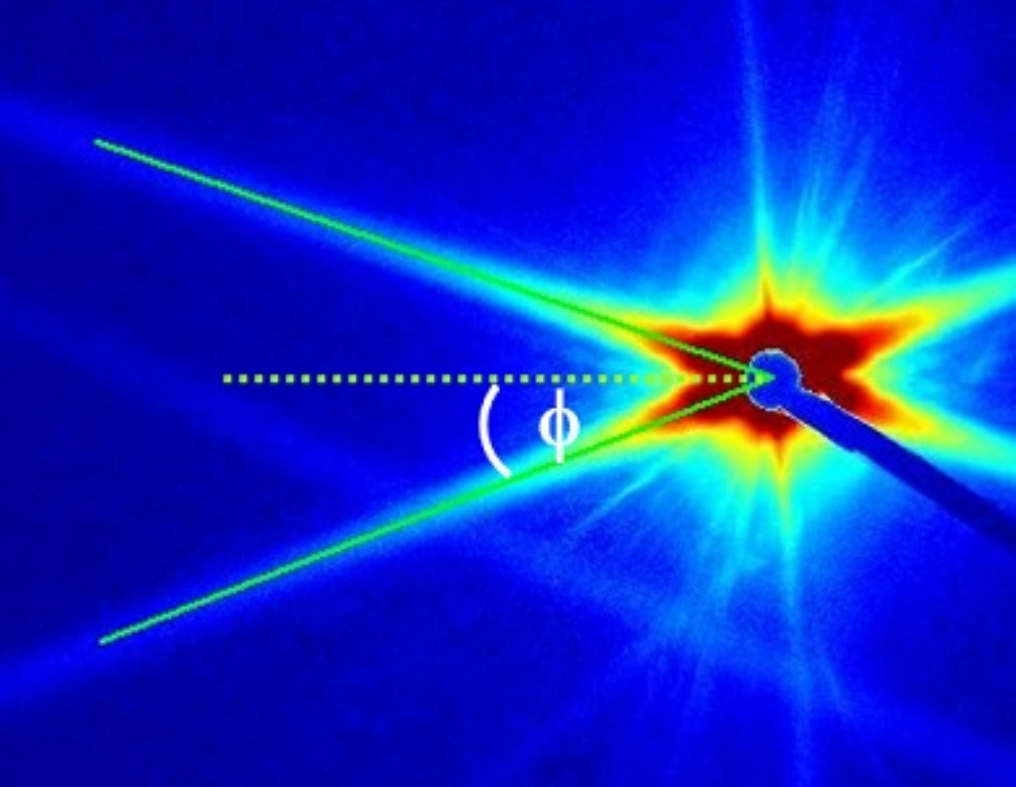}
 \caption{Example of measurement of the angle $\phi$ between the streaks of a scattering pattern.}
  \label{fig:Fig5}%
\end{figure}

Because the physical cones are three dimensional, there are two directions they could rotate with respect to the x-ray beam (Pitch and Yaw shown in Fig.~\ref{fig:Fig2}) which will affect the detected scattering image. Our model assumes radial symmetry about the etched cones. For etched ion tracks in amorphous SiO$_2$ this is a reasonable assumption which has been verified experimentally in our previous work \cite{Hadley2019} and in other studies (for example, see \cite{Dallanora2008, Jensen2006}). Rotation in the Roll direction does not have any effect on the scattering image when the cones are axially aligned with the x-ray beam. Tilting the sample in the direction of the Yaw angle $\xi$ will give identical results to tilting the sample by the Pitch angle $\gamma$, but the resulting scattering pattern will be rotated by 90\degree. This has been both simulated (see the attached supplementary information at URL XXXX) and verified experimentally. Because we have greater freedom of movement in the $\gamma$ direction (360\degree{} for $\gamma$ compared with about $\pm$ 12\degree{} for $\xi$), experimentally we choose to tilt the sample in the $\gamma$ direction. Choosing which angle to vary does not affect the subsequent analysis, but the effect of rotation of both angles together may be significant. No rotation is applied in the Roll direction once the sample is aligned. 

To accurately determine the value of $\beta$ it is important to be able to accurately determine the tilt angle since a potential source of error, albeit very small, may be introduced during the initial sample alignment with the x-ray beam. A rotation around any arbitrary axis in the XY plane can be decomposed into Yaw and Pitch rotations. Therefore, Equation~\ref{eqn:eqn18} can be reformulated to include both Yaw and Pitch. Provided $\xi$ is small, the decomposition of $\gamma$ into Pitch and Yaw can be approximated to $\frac{1}{2}(1+\mathrm{cos}\xi)\mathrm{sin}\gamma$ and substituted into Equation~\ref{eqn:eqn18} to yield Equation~\ref{eqn:eqn19}:

\begin{equation}
\mathrm{tan}\,\phi\, =\, \frac{2 \, \mathrm{tan} \,\beta}{(1 + \mathrm{cos} \,\xi_{offset}) \, \mathrm{sin} \,(\gamma + \gamma_{offset})}
\label{eqn:eqn19} \, .
 \end{equation}

A tilt sequence of measurements can be performed to determine any offset in the initial alignment. The $\gamma$ values may be plotted against the values of $\phi$ measured directly from the corresponding scattering images as shown for the example given in Fig.~\ref{fig:Fig5}. The fit of the plotted data may then be obtained using Equation~\ref{eqn:eqn19}. For the data presented the values of $\xi_{offset}$ were small and had no influence on the fit so it was set to zero. The fit of the plotted experimental data yields the cone half opening angle $\beta$, as well as any offset in the initial sample alignment in both possible directions ($\gamma_{offset}$ and $\xi_{offset}$). Note that the evaluation of the experimental images using Equation~\ref{eqn:eqn18}, and therefore Equation~\ref{eqn:eqn19}, is naturally restricted to images where streaks begin to appear, which is at values of $\gamma$ slightly larger than the cone angle (e.g. for the experimental data plotted in Fig.~\ref{fig:Fig7}, $\gamma \simeq$ 17-45\degree, and up to 90\degree{} for the simulations shown in the video).

%%%%%%%%%%%%%%%%%%%%%%%%%%%%%%%%%%%%%%%%%%%%%%%%%%%%%%%%%%%%%%%%%%%%%%%%%%%%%%%%%%%%%%%%%
\section{Results}
%%%%%%%%%%%%%%%%%%%%%%%%%%%%%%%%%%%%%%%%%%%%%%%%%%%%%%%%%%%%%%%%%%%%%%%%%%%%%%%%%%%%%%%%%

The analysis methods developed for accurate characterisation of conical nanopores were applied to two different types of samples. (i) An example representative of the series of data presented in \cite{Hadley2019}, with R $\approx$ 124 nm and H $\approx$ 550 nm, and (ii) a series of very high aspect ratio conical channels with very small opening angles ($<$ 3\degree{}) in 30 $\upmu$m thick polycarbonate film. Because the pores are etched completely through the PC foil, they resemble truncated cones, however this has no impact on the results.

\subsection{Conical pores in SiO$_2$}
%%%%%%%%%%%%%%%%%%%%%%%%%%%%%%%%%%%%%%%%%%%%%%%%%%%%%%%%%%%%%%%%%%%%%%%%%%%%%%%%%%%%%%%%%
Figure~\ref{fig:Fig4} shows an example of an experimental $\gamma$ tilt sequence for a representative sample irradiated with 89 MeV Au ions and etched in 2.5\% HF for 12 minutes. The scattering pattern obtained for the aligned image ($\gamma$ = 0\degree{}) shows a series of concentric rings, similar to a pinhole scattering experiment \cite{Guinier1955}. As the real space scattering objects are tilted by an angle $\gamma$ with respect to the x-ray beam, the scattering pattern in Fourier space starts to elongate until two opposite pairs of streaks begin to appear (at $\gamma \simeq$ 15\degree{} for this measurement sequence). As $\gamma$ is increased further, the angle between the arms of the streaks on the detector image begins to decrease. By substituting $\gamma$ = 90\degree{} into Equation \ref{eqn:eqn18} and assuming $\xi$ = 0\degree{} we can see that if we could rotate our sample to $\gamma$ = 90\degree{}, the angle between the arms of the streaks in the scattering pattern would be equal to the half cone opening angle, \cite{Guinier1955}, i.e. $\beta=\phi$.

The tilted scattering images contain information about the size and shape of the conical scattering objects. While the angle between the streaks is related to the cone opening angle, the oscillations in the streaks contain information about the cone radii. Figure \ref{fig:Fig6} shows two images selected from the tilt sequence in Fig.~\ref{fig:Fig4}, and their calculated fits obtained with our 2-D data analysis program. The fits reproduce the experimental images very well, with even the small secondary maxima between the streaks visible.

\begin{figure}
 \includegraphics[width=\figsize]%
 {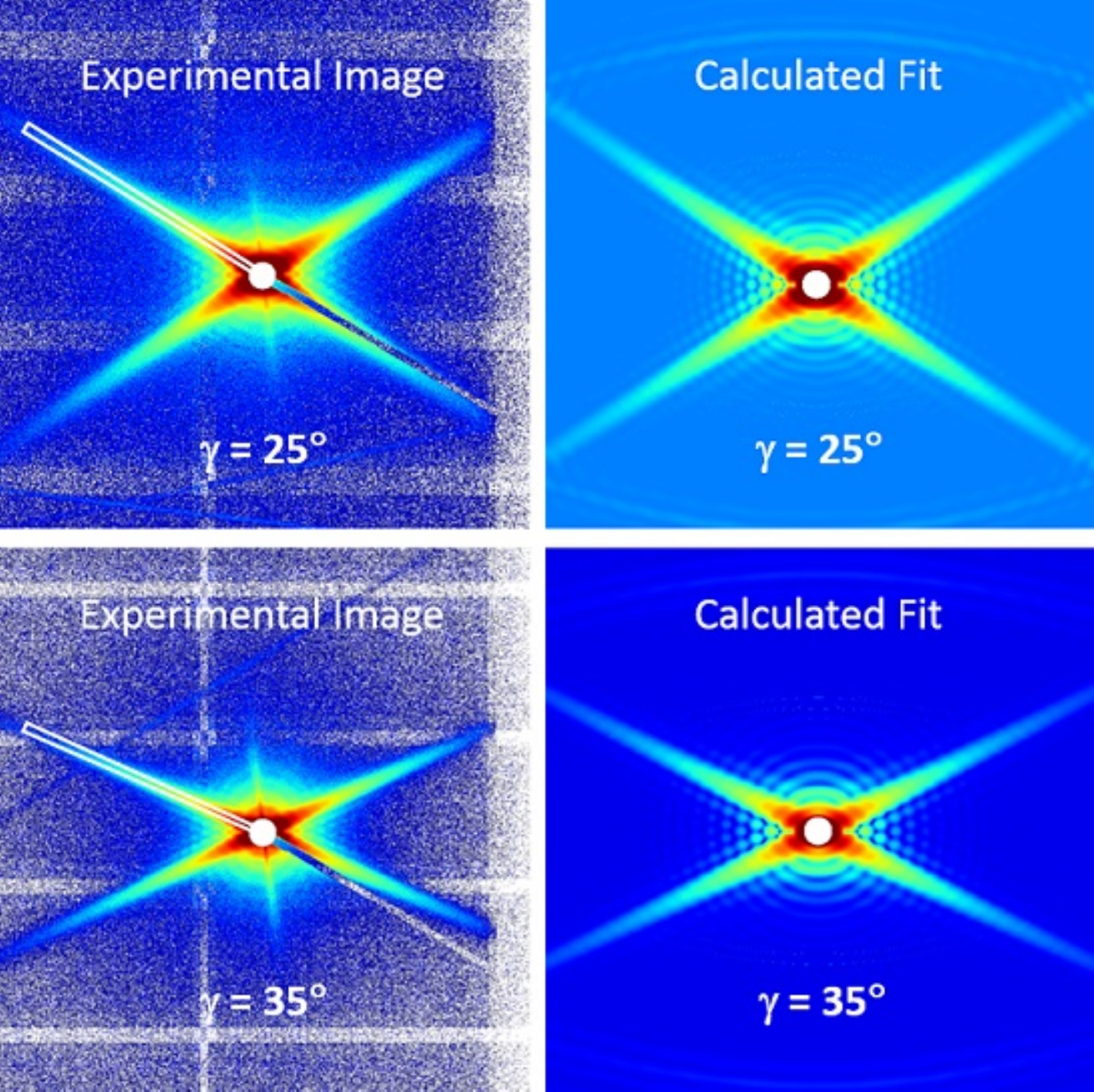}
 \caption{Example of a 2-D fit for two experimental images from the $\gamma$ tilt sequence shown in Fig.~\ref{fig:Fig4}. Masks used to extract 1-D data are indicated in white on the two experimental images on the left.}
  \label{fig:Fig6}%
\end{figure}

To accurately extract the parameters from the scattering images there are two options when analyzing the 2-D data. One approach is to fit each image from the whole $\gamma$ tilt sequence individually using our 2-D data model (using Equation~\ref{eqn:eqn10}) and average the results over the whole sequence of measurements. This was done for the data presented in \cite{Hadley2019}. The errors were obtained from averaging the measured values over the whole tilt sequence, rather than a single measurement, to provide a better estimate of the confidence in our measurements. An average over the tilt sequence allows for uncertainty introduced by factors like physical positioning in the x-ray beam, as well as the statistical variation in the fitting procedure.  For the example shown in Fig.~\ref{fig:Fig4} using this method the value obtained for $\beta$ was 12.69 $\pm$ 0.12\degree{}, with the base radius R$_0$ = 124 $\pm$ 5 nm and the cone height H = 551 $\pm$ 18 nm. This method works well provided the initial alignment of the sample with the x-ray beam is accurate and any initial misalignment is small. However, for samples with wide opening angles it can be difficult to determine how well they are aligned because the isotropic aligned pattern does not change significantly until the cones are tilted by an angle approaching the size of $\beta$, as can be seen by the first three tilted images in Fig.~\ref{fig:Fig4}.

The second approach enables us to determine whether there is an initial offset in the tilt angle $\gamma$. To determine the possible offset, we fit one image using an approximation for the value of $\gamma$. The best approximation is the nominal tilt angle relative to the aligned position in the experimental set-up. Because none of the parameters change apart from the tilt angle we can then fix all of the parameters except $\gamma$ and subsequently fit two images simultaneously. We also know the difference between the two $\gamma$ values for both images very accurately. This gives us the actual value of $\gamma$ including any offset angle due to the initial alignment, as well as any rotation offset in the orthogonal direction ($\xi_{offset}$). An increase in the value of $\xi$ has the effect of rotating the image. Two experimental images shown in Fig.~\ref{fig:Fig5}. were chosen from the tilt sequence from the SiO$_2$ sample shown in Fig.~\ref{fig:Fig4}, where $\gamma$ = 25\degree{} and 35\degree{}. The tilt angle difference is $\Delta\gamma$ = 10 $\pm$ 0.1\degree{}. The measured parameters obtained from the fit were $\beta$ = 12.82 $\pm$ 0.01\degree{}, $\gamma$ = 34.91 $\pm$ 0.02\degree{}, $\xi$ = 0.42 $\pm$ 0.01\degree{} and the cone height H = 549.8 $\pm$ 0.5 nm. The nominal value of $\gamma$ was 35\degree{}, so $\gamma_{offset}$ = 0.09\degree{}. Using Equation~\ref{eqn:eqn12} the base radius calculated is R$_0$ = 125.1 $\pm$ 0.5 nm. This compares well with the value of R$_0$ obtained with the first approach by averaging over the whole tilt sequence of images (R$_0$ = 124 $\pm$ 5 nm), however the uncertainty is improved by an order of magnitude suggesting the two image fit approach is more accurate when compared to the first approach of averaging over the whole tilt sequence.

To independently determine the value of $\beta$ from the scattering images of the $\gamma$ tilt sequence shown in Fig.~\ref{fig:Fig4}, the angle $\phi$ between the streaks in each of the scattering images was measured as shown in the example given in Fig.~\ref{fig:Fig5} using Fiji, an open source distribution of ImageJ \cite{Schindelin2012}. For clear streaks to appear the sample must be tilted by a critical angle slightly greater than the half cone opening angle $\beta$. In Fig.~\ref{fig:Fig4} streaks are beginning to appear when $\gamma$=15\degree, and are quite clear when $\gamma$=17\degree. The measured values of $\phi$ were then plotted for each corresponding value of $\gamma$ as shown in Fig.~\ref{fig:Fig7}. At lower tilt angles the streaks become broader which is reflected in the larger error estimates shown for $\phi$. Equation~\ref{eqn:eqn19} was used to fit the experimental data sequence, yielding the cone half opening angle $\beta$, the tilt angle offset $\gamma_{offset}$, and any offset in the orthogonal tilt angle $\xi_{offset}$. For the sample analysed in this example which were well aligned, $\xi_{offset}$ was set to zero. The measured values were $\beta$ = 12.87\degree{} $\pm$ 0.21\degree{} and $\gamma_{offset}$ = -2.03\degree{} $\pm$ 0.38\degree{}. The uncertainties given are the fitting uncertainty. For this sample the value obtained for $\beta$ agrees very well with the value obtained by the 2-D numerical fitting method.

\begin{figure}
 \includegraphics[width=\figsize]%
 {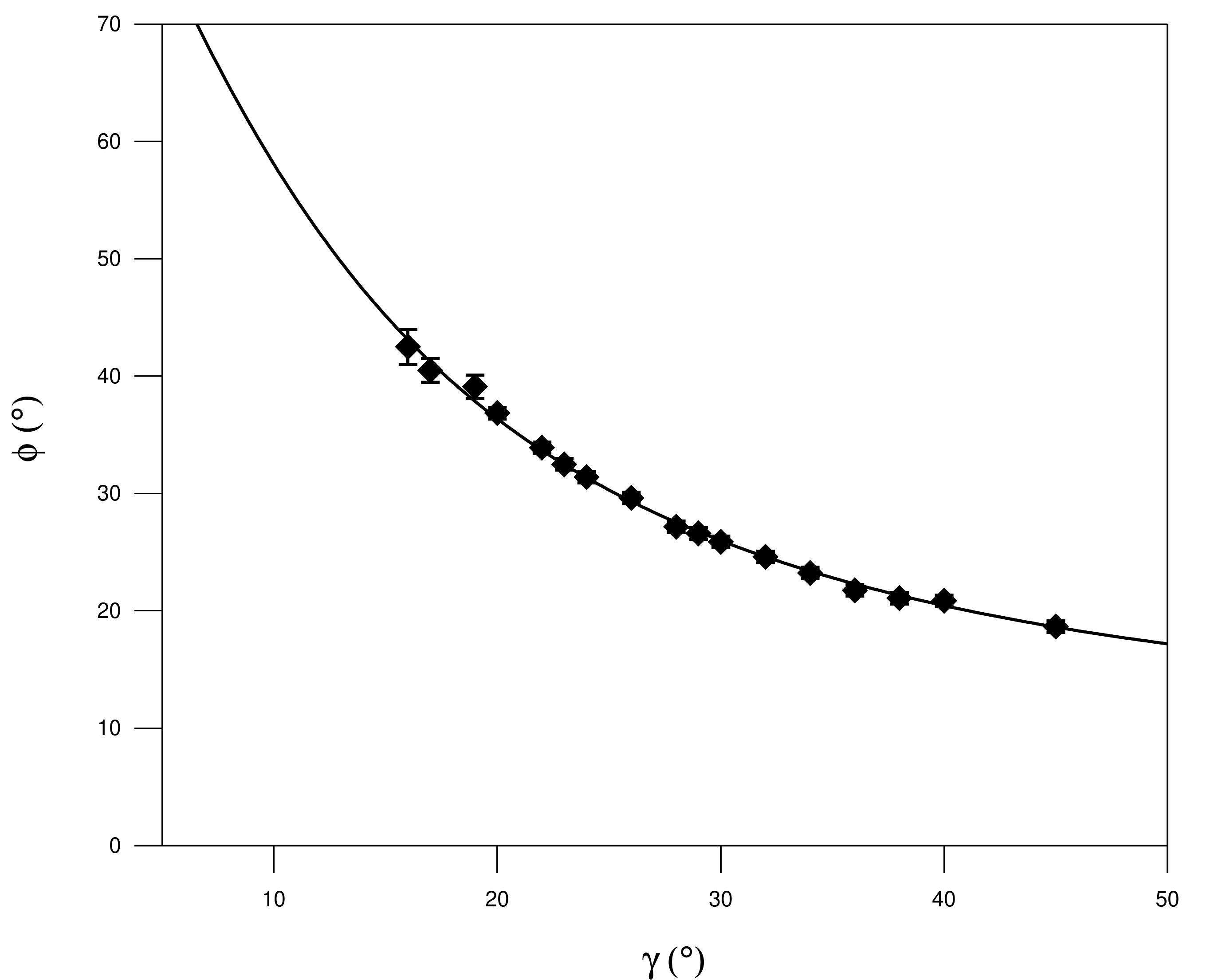}
 \caption{Plot of $\phi$ vs $\gamma$ for the $\gamma$ tilt sequence of images shown in Fig.~\ref{fig:Fig4}. The solid line is the fit obtained using Equation~\ref{eqn:eqn19}.}
  \label{fig:Fig7}%
\end{figure}

The uncertainty in the measured value of $\xi_{offset}$ could most likely be improved by performing a $\xi$ tilt sequence of measurements as well as the $\gamma$ tilt sequence, however the value of $\xi_{offset}$ is very small and very little improvement would be gained. Our experimental set-up did not allow us to perform a complete measurement tilt sequence in both the $\gamma$ and $\xi$ directions because of physical limitations of the movement of the goniometer (restricted to 12\degree{} in the Yaw direction). To visualise the effect of tilting the sample in both directions, we performed a simulation of the scattering intensity patterns as a function of the tilt angles. An example of the simulated effect of rotation of the sample with respect to the direction of the x-ray beam in both directions can be seen in the video, included as Supplemental Material at [URL XXXXX], where the movement of the real space cone is shown together with its corresponding scattering image.

Information about the size of the cones is contained in the oscillations along the streaks. Analysis of a 1-D data slice through a streak of the experimentally obtained image provides an alternative means for determining the pore dimensions. The 1-D data fit is computationally much easier, which can be useful in some circumstances. A narrow mask (indicated by the white rectangle) is used to extract the data, as shown for the two examples of experimental images on the left of Fig. \ref{fig:Fig6}. An air shot was used for background subtraction, and the intensity oscillations along the streaks become apparent when the data is plotted as a function of the scattering vector (as shown in Fig.~\ref{fig:Fig8}), for selected tilt angles from the $\gamma$ tilt sequence shown in Fig.~\ref{fig:Fig4}. To fit the data plotted in Fig.~\ref{fig:Fig8}, the hard cylinder model was used as an approximation for the cones with Equation~\ref{eqn:eqn13} representing the scattering amplitude and using the square distribution coherently summed (Equation~\ref{eqn:eqn17}), the model effectively fits a series of cylinders with radii corresponding to the cone radii varying as a function of $\beta$ along the cone's length. The reason why the data for $\gamma$=0\degree{} was not fit is that accurate alignment of the pores with the x-ray beam is required to yield good 1-D data when integrated. Any misalignment skews the data. The data and fits plotted in Fig.~\ref{fig:Fig8} show that this model provides a very good fit for the tilted samples.

\begin{figure}
 \includegraphics[width=\figsize]%
 {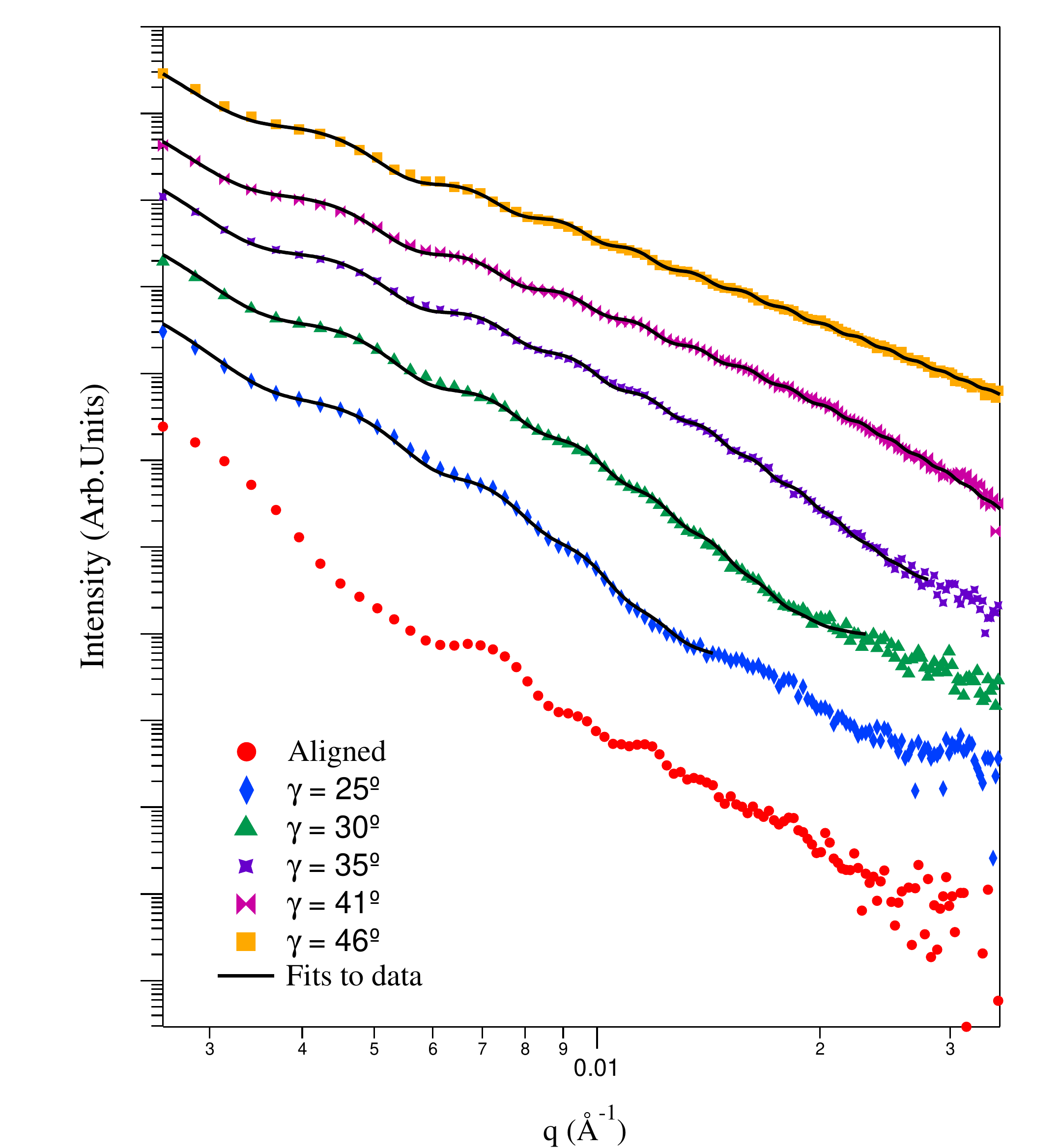}
 \caption{1-D data plots from selected images from the $\gamma$ tilt sequence shown in Fig.~\ref{fig:Fig4}. The fits to the data were obtained using Equation 13 and a box shaped distribution for the radius variation (Equation 17). To enable comparison, the data are offset on the intensity axis for clarity.}
  \label{fig:Fig8}%
\end{figure}

The values obtained for the cone radius and distribution width from the data plotted in Fig.~\ref{fig:Fig8} for each tilt angle are given in Table~\ref{tab:table1}. The radius values measured correspond to the volume weighted average radius (R$^{\prime}$) of the cones, which is the value of the radius at $1/4$ of the height (or centroid) of the cone. To check this is correct we can compare the experimental results for the radius obtained by fitting the 1-D data (R$^{\prime}$ = 92.2 $\pm$ 1.0 nm) with the radius value obtained from the 2-D data fit (R$_0$ = 124 $\pm$ 5 nm, or R$_{0}$ = 125.1 $\pm$ 0.5 nm in the case of the two image simultaneous fit). In both cases the average radius values over the tilt sequence was used. The value of $\beta$ obtained by fitting the tilt sequence of Fig. 4 using the 2-D fitting procedure was 12.97 $\pm$ 0.07\degree{}. Because we have measured the cone angle to a very high degree of certainty, we can use Equation 11 to confirm that 92.2 $\pm$ 1.0 nm corresponds to the radius value at the cone centroid, with a calculated base cone radius of R$_{0}$ = 123 $\pm$ 1 nm. This compares very well with the base cone radius measured by the 2-D fit which was R$_{0}$ = 124 $\pm$ 5 nm, or R$_{0}$ = 125.1 $\pm$ 0.5 nm for the two image fit. These results are summarised in Table~\ref{tab:table2}.

\begin{table}[h]
%\texttt{table}
\caption{Parameters extracted from 1-D fits plotted in Fig.~\ref{fig:Fig8}.}
\label{tab:table1}
%\centering
\begin{ruledtabular}
\begin{tabular}{l l l}
$\gamma$(\degree) & R$^{\prime}$ (nm) & W (nm) \\
\hline
25 & 89.6$\pm$0.1 & 38.7$\pm$4.4 \\
30 & 91.2$\pm$1.6 & 40.0$\pm$1.6 \\
35 & 94.5$\pm$1.1 & 38.6$\pm$0.9 \\
41 & 92.5$\pm$0.1 & 45.3$\pm$1.5 \\
46 & 93.3$\pm$1.2 & 45.6$\pm$1.2 \\
\hline
Average radius = & 92.2$\pm$1.0 &  \\
\end{tabular}
\end{ruledtabular}
\end{table}

\begin{table}[h]
\caption{Comparison of results obtained by each analysis method.}
\label{tab:table2}
\begin{ruledtabular}
\begin{tabular}{l l l l l}
Parameter & 2-D fit & 2-D 2 image fit & Angle fit function & 1-D fit \\
\hline
$\beta$(\degree) & 12.69$\pm$0.12 & 12.82$\pm$0.01 & 12.87$\pm$0.21 & 12.82$\pm$0.01 \\
$R_0$ (nm) & 124$\pm$5 & 125.1$\pm$0.5 &  & 123$\pm$1 \\
H (nm) & 551$\pm$18 & 549.8$\pm$0.5 &  & 540$\pm$1 \\
$\gamma$ (\degree) & 35 (nominal) & 34.91$\pm$0.02 &  \\
$\gamma_{offset}$ (\degree) & & 0.09$\pm$0.02 & -1.76$\pm$0.07 & \\
$\xi_{offset}$ (\degree) & & 0.42$\pm$0.01 & 0.7$\pm$13.6 & \\
R$^{\prime}$ (nm) &  &  &  & 92$\pm$1  \\
\end{tabular}
\end{ruledtabular}
\end{table}

\subsection{High aspect ratio etched cones in polycarbonate}
%%%%%%%%%%%%%%%%%%%%%%%%%%%%%%%%%%%%%%%%%%%%%%%%%%%%%%%%%%%%%%%%%%%%%%%%%%%%%%%%%%%%%%%%%

Figure~\ref{fig:Fig9} shows scattering images from SAXS data obtained from a series of approximately 30 $\upmu$m long conical etched ion tracks in polycarbonate (PC) foil. A $\gamma$ tilt sequence of scatting intensity images for a sample etched in 9M NaOH with 30\% methanol added is shown in Fig.~\ref{fig:Fig9}(a). Figure~\ref{fig:Fig9}(b) shows the scattering images for the whole series of six samples etched with varying methanol concentration (including 0$\%$) all tilted by the same amount ($\gamma$ = 10\degree{}). The effect of varying the methanol concentration in the etchant is clearly apparent from the difference of the angles between the streaks (Fig.~\ref{fig:Fig9}(b)).

\begin{figure}
 \includegraphics[width=\figsizeb]%
 {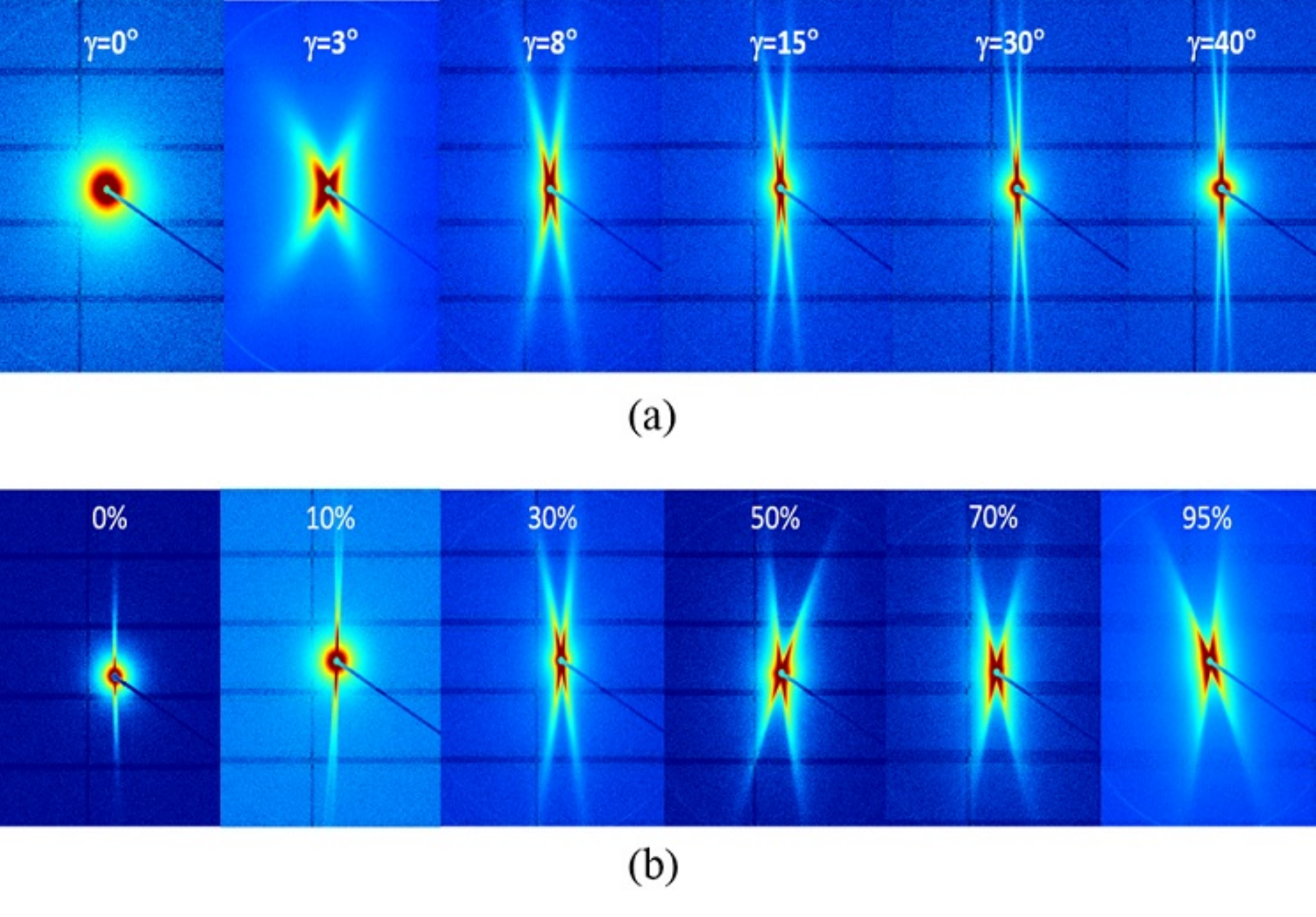}
 \caption{Scattering intensity images for (a) $\gamma$ tilt sequence for conical etched ion tracks in PC with 30\% methanol added and (b) six samples etched using different methanol concentrations measured at a constant tilt angle ($\gamma$ = 10\degree{}).}
  \label{fig:Fig9}%
\end{figure}

Figure~\ref{fig:Fig10} compares the 1-D data extracted from the streak of a sample etched with 0\% methanol and 30\% methanol added to the etchant both tilted by $\gamma$ = 10\degree{}. The sample etched with 0\% methanol is expected to have practically cylindrical pores. This is reflected in the scattering image since there is only one pair of streaks. The plotted 1-D data shows some intensity oscillations. Assuming cylindrical geometry, the hard cylinder model yielded a radius value of 54.7 $\pm$ 2.4 nm. In contrast, the 1-D data plotted for the sample with conical channels shows no oscillations, so fitting using the cylinder model will yield results with large uncertainties. Although just one example of the 1-D data was plotted in Fig.~\ref{fig:Fig11}, the streaks for all the conical samples are similar as they also do not show any oscillations. The reason why there are no oscillations for the conical samples is because the detected scattering intensity is the sum of the scattering over the entire length of these very long cones. The length (~30 $\upmu$m) by far exceeds the coherence length ($\sim$500 nm) of the x-ray beam, effectively incoherently summing contributions from pore segments with a large size distribution. This flattens out the oscillations that are characteristic in scattering from monodisperse samples. For almost cylindrical channels (0\% methanol) there is only a very small variation in the radius of the cones along their length, so the oscillations are still apparent.

\begin{figure}
 \includegraphics[width=\figsize]%
 {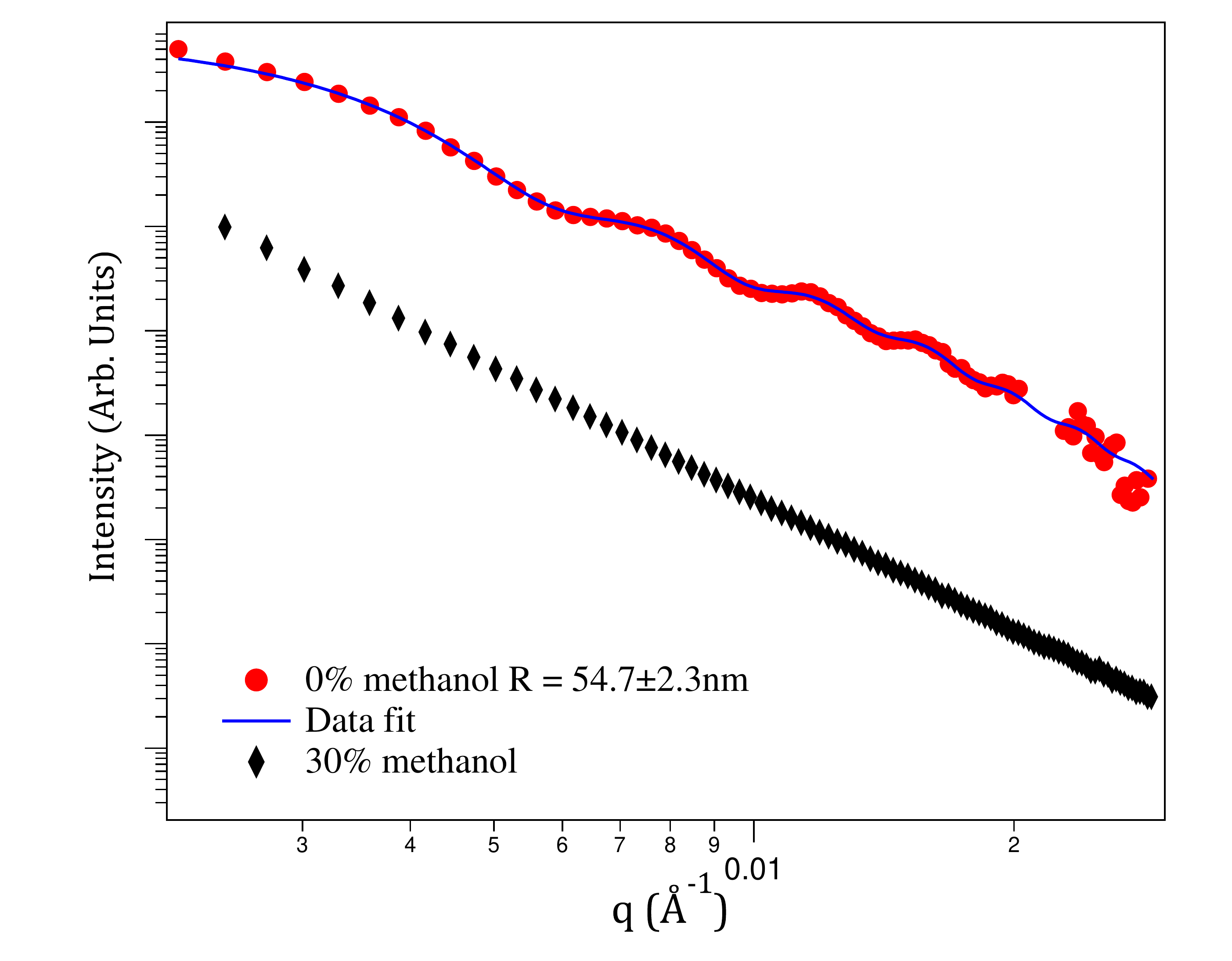}
 \caption{1-D data comparison for an almost cylindrical sample etched in 9M NaOH with no methanol, compared with conical pores etched with 30\% methanol showing the effect of the conical pore shape on the oscillations in the data.}
  \label{fig:Fig10}%
\end{figure}

Figure~\ref{fig:Fig11} is a plot of the angle $\phi$ between the streaks measured directly from the scattering intensity patterns for each sample, plotted against the tilt angle $\gamma$. The solid lines are fits to the data obtained using Equation~\ref{eqn:eqn19}. The values of $\beta$ deduced from the fits are given in Table~\ref{tab:table3}. Although it is not possible to make a direct comparison because the etching conditions are not exactly the same, the results obtained for $\beta$ from the data fits suggest the etched ion tracks have a similar shape to the Cu cone replicas produced by electrodeposition in ion track membranes etched under similar conditions and analysed by electron microscopy in \cite{Duan2016}. The method used in our work takes advantage of superior statistics, since a single SAXS measurement at a fluence of 10$^{6}$ ions cm$^{-2}$ simultaneously measures around 2 x 10$^{4}$ etched tracks. The effect of increasing methanol concentration on the cone opening angle can be seen in the plot shown in Fig. \ref{fig:Fig12}, where the solid line shows the data trend. The addition of methanol affects the ratio of the ion track etch rate to the bulk etch rate ($\nu_{t}/\nu_{b}$). The higher the methanol concentration, the lower this ratio resulting in a larger value of $\beta$.

\begin{figure}
 \includegraphics[width=\figsize]
 {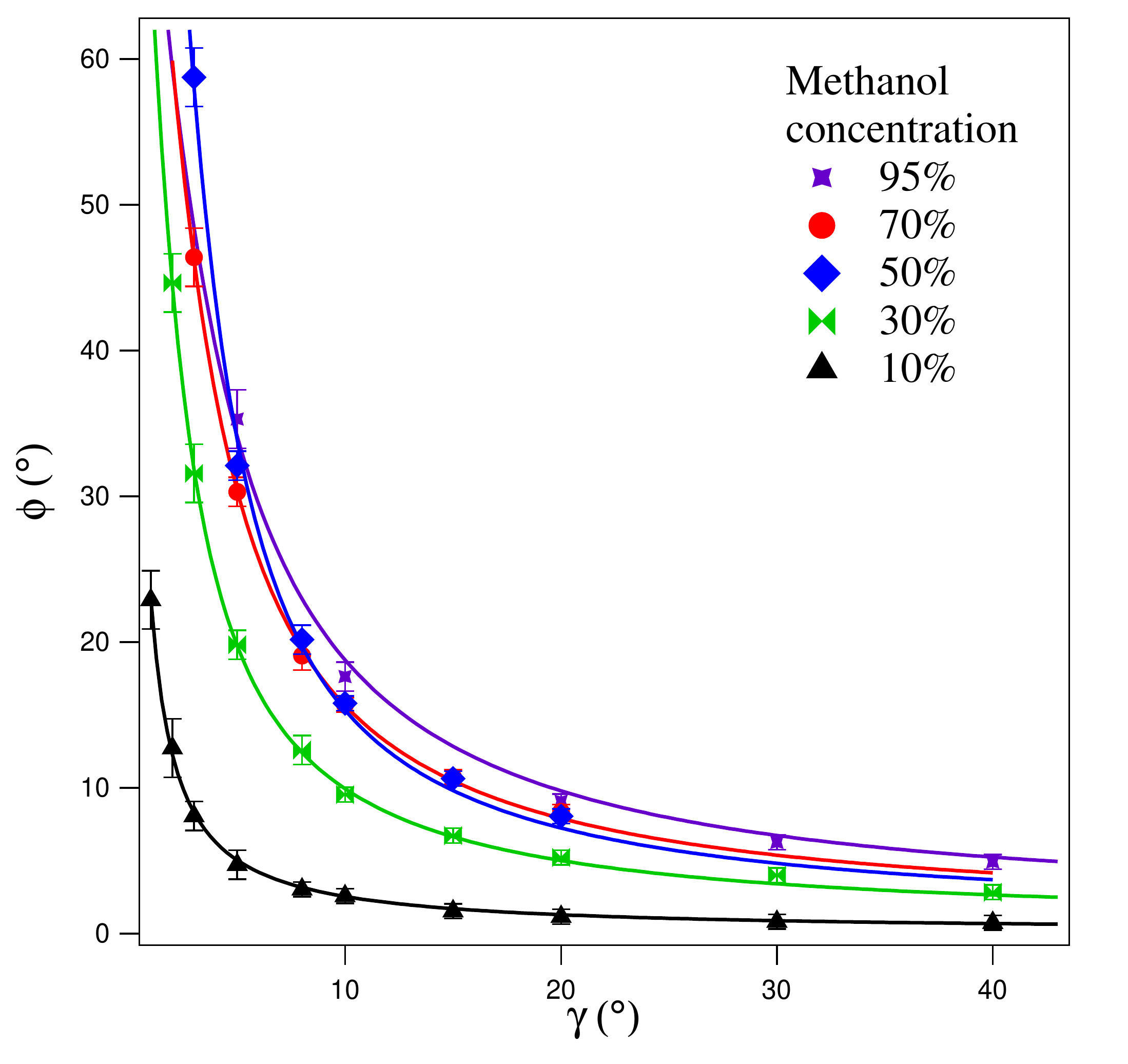}
 \caption{Plot of $\phi$ against $\gamma$ for the series of samples etched with varying concentrations of methanol. The solid lines are fits to the data using Equation 19 which yield the cone half opening angle $\beta$.}
  \label{fig:Fig11}
\end{figure}

\begin{figure}
 \includegraphics[width=\figsize]
 {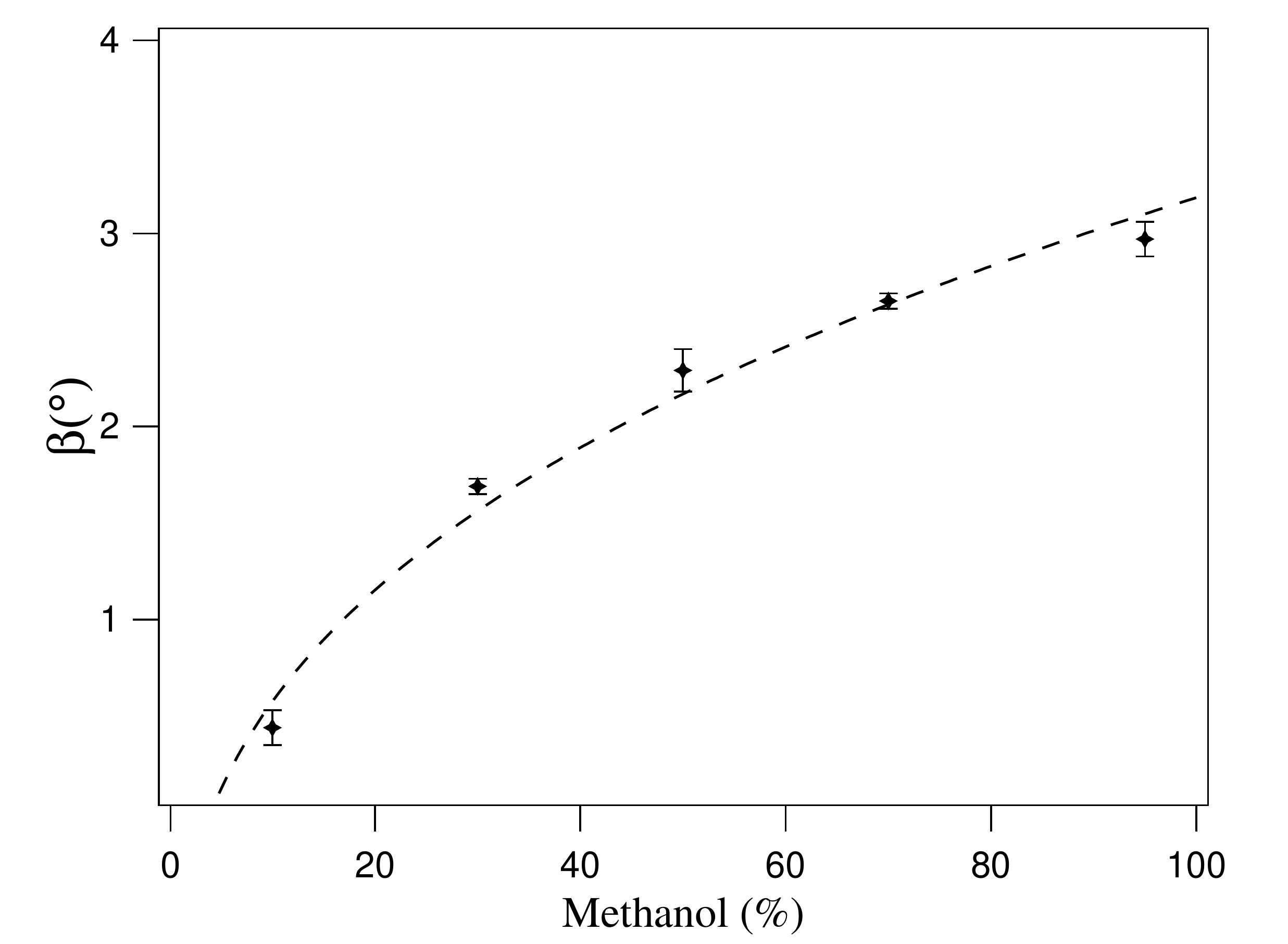}
 \caption{Plot of $\beta$ against etchant methanol concentration showing the increase in the cone opening angle with increased methanol concentration. The dashed line is to show the trend in the data.}
  \label{fig:Fig12}
\end{figure}

\begin{table}[h]
\caption{Half cone opening angles ($\beta$) deduced from fits to the data plotted in Fig. \ref{fig:Fig11}.}
\label{tab:table3}
\begin{ruledtabular}
\begin{tabular}{l l l}
Methanol (\%) & Etching time (min) & $\beta$(\degree) \\
\hline
0 & 7 & N/A \\
10 & 7.5 & 0.44$\pm$0.09 \\
30 & 8.5 & 1.69$\pm$0.04 \\
50 & 11 & 2.29$\pm$0.11 \\
70 & 19.5 & 2.65$\pm$0.04 \\
95 & 80 (approx) & 2.96$\pm$0.06
\end{tabular}
\end{ruledtabular}
\end{table}

\section{conclusions}

In this paper we present new techniques for the analysis of SAXS results obtained from conical mono-disperse nanochannels. These techniques are well suited to accurately extract parameters for conical etched ion tracks in a-SiO$_2$ as well as for very high aspect ratio conical etched ion tracks in polycarbonate. The two material systems studied were very different from the perspective of the aspect ratios of the conical etched channels in each, demonstrating that the analysis techniques would be applicable to pores in many other materials.

The 2-D fitting routine we developed enables us to measure the axial and radial track etch rates and the cone opening angles with high precision. The 2-D fits reproduce the experimental images very well, with even the very fine secondary features between the streaks visible. By analysing a tilt sequence of SAXS measurements, we were able to determine if there was any tilt angle offset introduced by the experimental set-up, making sure that the deduced values of the cone opening angles were accurate to better than $\pm$ 0.1\degree.

As well as the 2-D fitting, we developed an independent technique to measure the cone opening angles from a tilt sequence of measurements. By exploiting the relationship between the cone opening angle and the angle between the streaks in the scattering pattern, we can determine the cone opening angle very accurately directly from the scattering images without needing to analyse the intensities. From the measurement tilt sequence using this method we can also measure and allow for any offset in the tilt angle, providing very high confidence in the measured values for the cone opening angles. Even if the scattering image was weak or there were no oscillations present in the streaks, we were still able to determine $\beta$ with very high confidence directly from the scattering intensity images.

Fitting a slice of the 1-D data through the streak using the hard cylinder model provides an alternative method for analyzing the data. With this method we could deduce the average cone radius and compare it with the radius values obtained with the 2-D fit. This method was useful for providing a good starting point for estimates of parameters for the 2-D fit, ensuring it returned accurate results and is also well suited for automated quick analysis given the short fitting times.

\begin{acknowledgments}

\section{Acknowledgements}

Part of the research was undertaken at the SAXS/WAXS beamline at the Australian Synchrotron, part of ANSTO, and we thank the beamline scientists for their technical assistance. Irradiation was conducted at the ANU Heavy Ion Accelerator Facility (HIAF), and we also thank the ANU HIAF staff for their technical assistance. Operations of the ANU HIAF is financially supported by the National Collaborative Research Infrastructure Strategy (NCRIS) HIA capability. The results presented here are based in part on UMAT experiments, which were performed at the X0-beamline of the UNILAC at the GSI Helmholtzzentrum fuer Schwerionenforschung, Darmstadt (Germany), in the frame of FAIR Phase-0. This work was also performed in part at the Canberra node of the Australian National Fabrication Facility, a company established under the NCRIS to provide nano and micro-fabrication facilities for Australia's researchers. This research is supported by an Australian Government Research Training Program (RTP) Scholarship. The authors also acknowledge financial support from the Australian Research Council (ARC) and the German Research Foundation (DFG).

A.H. irradiated and etched the SiO$_2$ samples, conducted the SAXS measurements and simulations, analyzed the SAXS data, and drafted the manuscript. C.N. wrote the 2-D SAXS data analysis software, assisted with the SiO$_2$ irradiation and SAXS measurements and reviewed the manuscript. P.M.S. assisted with the SiO$_2$ irradiation and SAXS measurements and reviewed the manuscript. S.D. generated the 3-D cone images, assisted with the SiO$_2$ irradiation and SAXS measurements. M.A.C.S. assisted with etching the PC samples, M.E.T.M. and C.T. irradiated the PC samples and reviewed the manuscript. S.M. assisted with the SAXS measurements. P.K. assisted with the SAXS measurements and data analysis, etched the PC samples and helped draft the manuscript. P.K. conceived and directed the research.

\end{acknowledgments}

% Create the reference section using BibTeX:
\bibliography{Cone_SAXS_AH}
%\bibliography{MyLibrary}

\end{document}